\documentclass[runningheads]{llncs}
\usepackage{graphicx}
\usepackage{amsmath}
\usepackage{amssymb}
\usepackage{url}
\usepackage{float}
\begin{document}
\title{Exploring Large Context for Cerebral Aneurysm Segmentation}
%
%
\author{Jun Ma\inst{1} \and Ziwei Nie\inst{2}}
\institute{Department of Mathematics, Nanjing University of Science and Technology \email{junma@njust.edu.cn} \and
Department of Mathematics, Nanjing University}
\authorrunning{J. Ma.}

%
%
\maketitle              
\begin{abstract}
Automated segmentation of aneurysms from 3D CT is important for the diagnosis, monitoring, and treatment planning of the cerebral aneurysm disease. This short paper briefly presents the main technique details of the aneurysm segmentation method in MICCAI 2020 CADA challenge. The main contribution is that we configure the 3D U-Net with a large patch size, which can obtain the large context. Our method ranked second on the MICCAI 2020 CADA testing dataset with an average Jaccard of 0.7593.
Our code and trained models are publicly available at \url{https://github.com/JunMa11/CADA2020}.

\keywords{Segmentation \and Deep learning \and U-Net \and CT.}
\end{abstract}
\section{Introduction}
Accurate segmentation of aneurysms plays an important role in the quantitative analysis for risk assessment and monitoring of the cerebral aneurysm disease. Manual segmentation is time-consuming and suffers from inter- and intra-observer variability. Thus, fully automatic aneurysm segmentation methods are highly demanded.
In MICCAI 2020, Cerebral Aneurysm Segmentation Challenge was hold to benchmark different segmentation methods. The main goal of this short paper is to present the main technique details of our methods. Thus, the medical background of aneurysms and the overview of the state-of-the-art aneurysm segmentation methods are out of the scope of this paper. More details of the challenge background and motivation are available in the challenge design document (\url{http://doi.org/10.5281/zenodo.3715011}).

\begin{figure}
\centering
\includegraphics[scale=0.6]{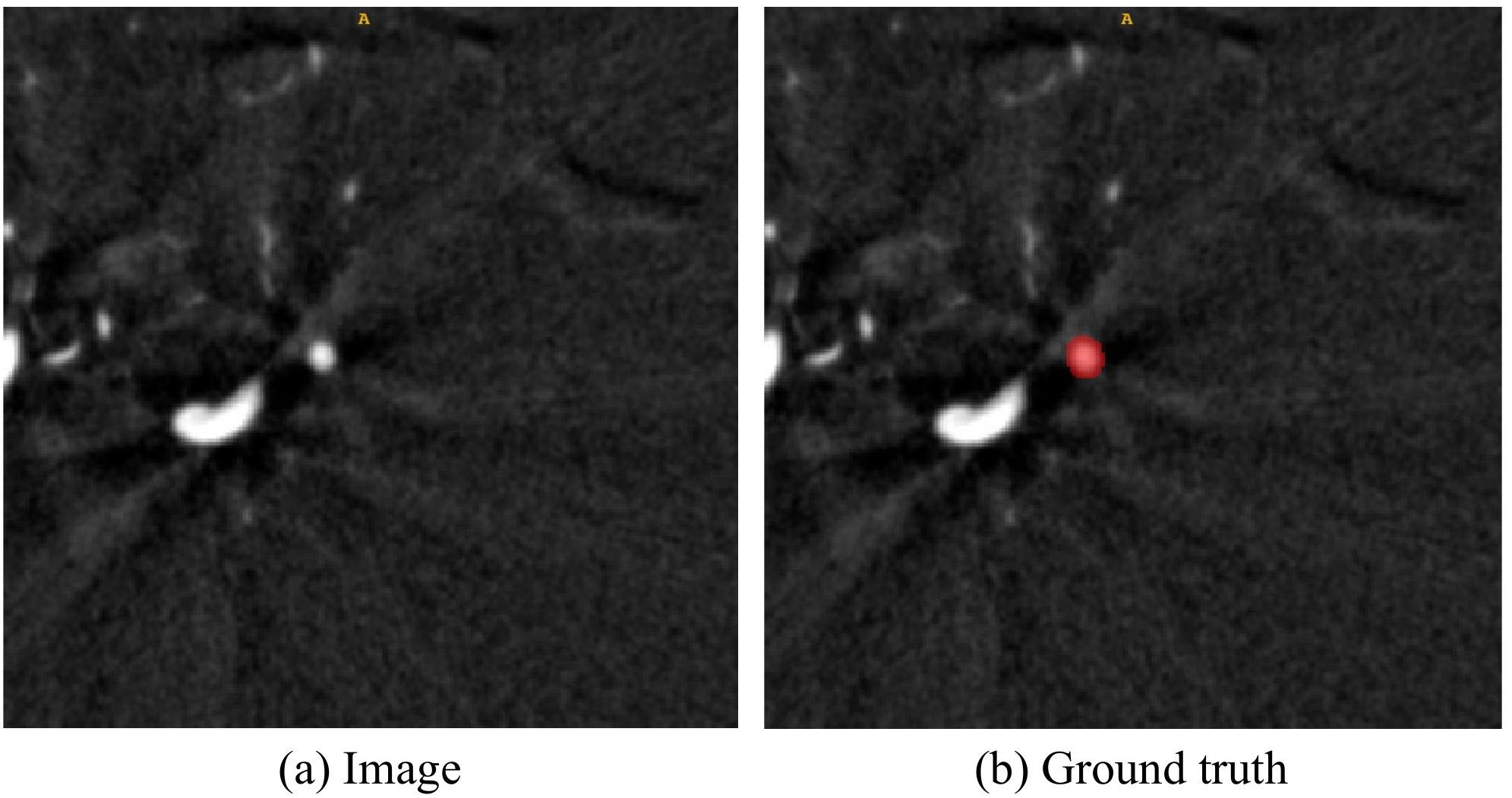}
\caption{An example of the aneurysm CT image and the corresponding ground truth.} \label{fig:example}
\end{figure}

Automatic segmentation of aneurysms is a challenging problem. Figure~\ref{fig:example} presents an example of the aneurysm CT image and the corresponding ground truth. It can be found that
\begin{itemize}
    \item the aneurysm is very small that only occupies very small proportion in the whole image.
    \item some other tissues share the similar appearances with the aneurysm, which are hard to distinguish them.
\end{itemize}
Our main motivation was from the first-place solution \cite{BraTS2018-1st} in the MICCAI 2018 brain tumor segmentation challenge (BraTS) where using a large image patch size  (i.e., $160\times192\times128$) can obtain better performance than a smaller image patch size (even with batch normalization). The potential reason might be that large patch size can provide more semantic context information for the network.

\section{Dataset and Method}
The challenge organizers provided 110 cases for training and 22 cases for testing. Specifically, image data of patients with cerebral aneurysms without vasospasm were collected for diagnostic and treatment decision purposes.
The image data were acquired utilizing the digital subtraction AXIOM Artis C-arm system using a rotational acquisition time of 5s with 126 frames. Post-processing was performed using LEONARDO InSpace 3D (Siemens, Forchheim, Germany). A contrast agent (Imeron 300, Bracco Imaging Deutschland GmbH, Germany) was manually injected into the internal carotid (anterior aneurysms) or vertebral (posterior aneurysms) artery. Reconstruction of a volume of interest selected by a neurosurgeon generated a stack of 220 image slices with matrices of 256x256 voxels in-plane, resulting in an iso-voxel size of 0.5 mm.

In our solution, 3D nnU-Net~\cite{UNet3d,nnunet2020} was used as the main network architecture. In particular, we modified the default nnU-Net with a large image patch size of $192\times224\times192$ and a batch size of 2 as the input, which were the largest patch size and batch size where the GPU memory allowed.
The detailed settings are as follows:
\begin{itemize}
    \item Preprocessing. We use three-order interpolation to resample all the images to a common spacing of $0.5429\times0.5429\times0.5429 mm^3$, and normalize the intensity to a mean of 0 and standard deviation of 1.
    \item Training. The U-Net has six resolutions. The feature size is decreased a half in each resolution via downsampling with strided convolutions. The optimizer is stochastic gradient descent with an initial learning rate (0.01) and a nesterov momentum (0.99). To avoid overfitting, standard data augmentation techniques are used during training, such as rotation, scaling, adding Gaussian Noise, gamma correction. The loss function is the unweighted sum of Dice loss and cross entropy. We apply five-fold cross validation with 110 training cases. Each fold is trained on a NVIDIA TITAN V100 GPU.
    \item Inference. The 5 models resulting from training are used as an ensemble for predicting the test cases.
\end{itemize}

\section{Results}
\begin{figure}[!htbp]
\centering
\includegraphics[scale=0.5]{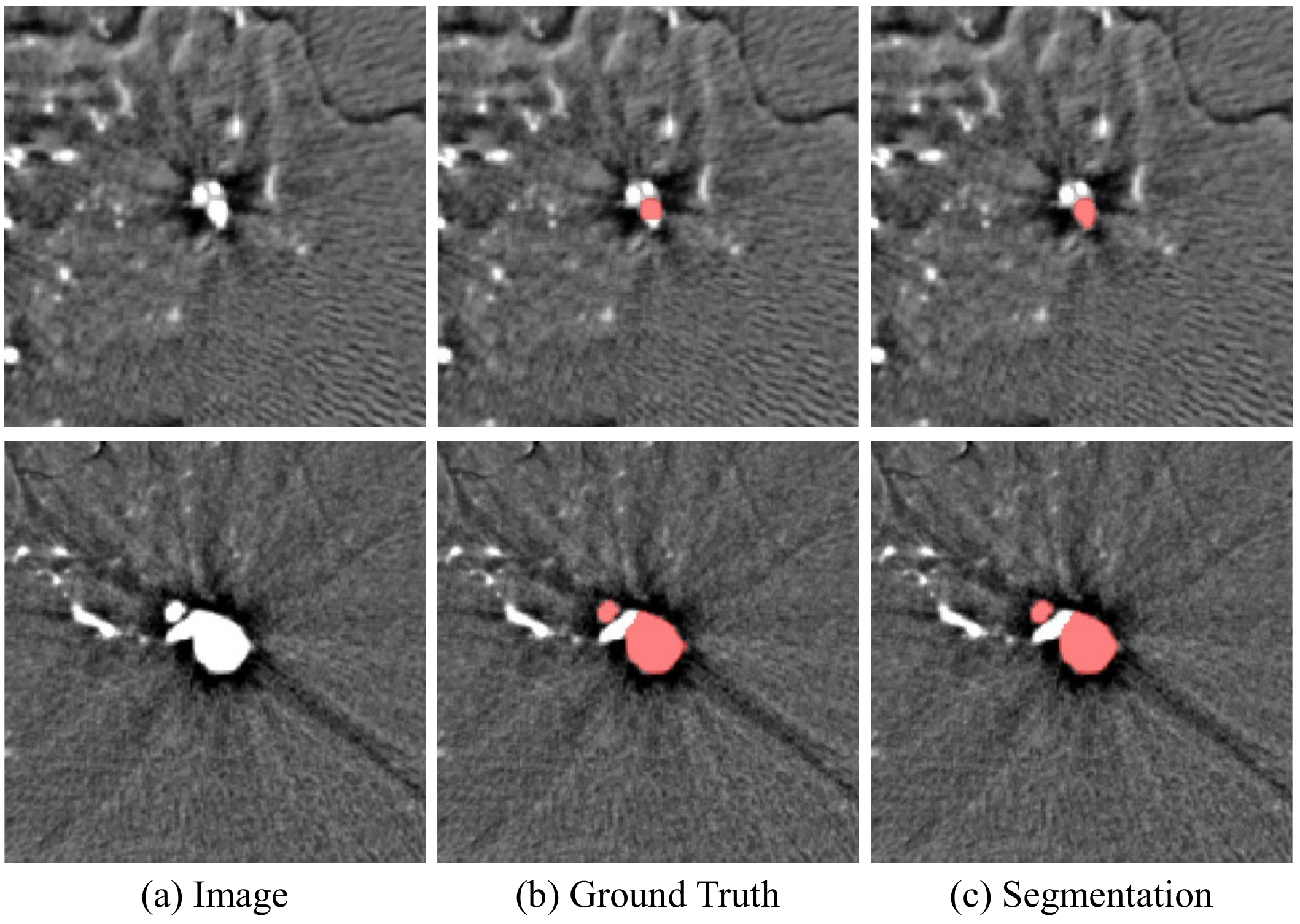}
\caption{Visualized examples in the cross-validation segmentation results.} \label{fig:cvseg}
\end{figure}

Table~\ref{tab:5fold} shows the quantitative results of the five-fold cross-validation.
The performances vary among different folds, indicating that the training cases have different difficulties.
Our method achieves an average Jaccard of 0.8112, Dice of 0.8861, precision of 0.8934 and recall of 0.9036 in cross validation.
Figure~\ref{fig:cvseg} presents some visualized segmentation results. Overall, the segmentation results are accurate. However, small segmentation errors can significantly degenerate the Jaccard/Dice scores because the aneurysms are very small.

\begin{table}[!htbp]
\caption{Quantitative results of five-fold cross-validation results on the training set. AVG denotes the average results of all training cases.}\label{tab:5fold}
\centering
\begin{tabular}{ccccc}
\hline
Fold & Jaccard & Dice   & Precision & Recall \\ \hline
0    & 0.7901                         & 0.8737 & 0.8970    & 0.8742 \\
1    & 0.8335                         & 0.9034 & 0.9046    & 0.9173 \\
2    & 0.7966                         & 0.8805 & 0.8611    & 0.9163 \\
3    & 0.7718                         & 0.8470 & 0.8661    & 0.8904 \\
4    & 0.8638                         & 0.9256 & 0.9384    & 0.9197 \\ \hline
AVG  & 0.8112                         & 0.8861 & 0.8934    & 0.9036 \\ \hline
\end{tabular}
\end{table}

We apply the proposed method on the hidden testing set where the ground truth are held by the organizers. The total reference time is 88 minutes. Table~\ref{tab:testing} presents the quantitative segmentation results. It can be found that the average Jaccard score is 0.7593 that is lower than the cross validation results. We noticed that the minimal Jaccard score is 0, indicating that there are some very challenging cases that our method fails to handle.

\begin{table}[!htbp]
\caption{Quantitative results on the testing set.}\label{tab:testing}
\centering
\begin{tabular}{lc}
\hline
 Metric         & Value \\ \hline
Jaccard                       & 0.759                        \\
Volume Bias                   & 75.8                    \\
Mean Distance                 & 3.54                         \\
Volume Pearson R              & 0.998                        \\
Hausdorff Distance            & 4.97                         \\ \hline
\end{tabular}
\end{table}

\section{Conclusion}
In this short paper, we present our segmentation method for cerebral aneurysm segmentation. Specifically, we employed the well-known 3D U-Net and configured with a large patch size. This simple method achieved the second place with a Jaccard score of 0.7593 on the testing set (\url{https://cada-as.grand-challenge.org/FinalRanking/}). In future, we will speed up the inference time to make our method more efficient.

\section*{Acknowledgement}
This work is supported by the National Natural Science Foundation of China (No. 11531005, No. 11971229).
We are grateful to the High Performance Computing Center of Nanjing University for supporting the blade cluster system to run the experiments.
We also highly appreciate the CADA organizers for holding the great challenge and creating the publicly available dataset.
%
%

\bibliographystyle{splncs04}
\bibliography{REF}

\end{document}